\documentclass[aps,prl,10pt,twocolumn,superscriptaddress,longbibliography]{revtex4}
\usepackage{graphicx}
\usepackage{amssymb, amsmath}
\usepackage{color}
\usepackage{ulem}
\usepackage{bm}
\usepackage{graphicx}
\usepackage{subfigure}
\usepackage{dcolumn}
\usepackage{mathtools}
\usepackage{pifont}

\begin{document}

\title{Emergence of new van Hove singularities in the charge density wave state of a topological kagome metal RbV$_3$Sb$_5$}

\author{Soohyun Cho}
\thanks{Equal contributions}
\affiliation{Center for Excellence in Superconducting Electronics, State Key Laboratory of Functional Materials for Informatics, Shanghai Institute of Microsystem and Information Technology, Chinese Academy of Sciences, Shanghai 200050, China}

\author{Haiyang Ma}
\thanks{Equal contributions}
\affiliation{School of Physical Science and Technology, ShanghaiTech University, Shanghai 201210, China}
\affiliation{Key Laboratory of Artificial Structures and Quantum Control (Ministry of Education), Shenyang National Laboratory for Materials Science, School of Physics and Astronomy, Shanghai Jiao Tong University, Shanghai 200240, China}

\author{Wei Xia}
\thanks{Equal contributions}
\affiliation{School of Physical Science and Technology, ShanghaiTech University, Shanghai 201210, China}
\affiliation{ShanghaiTech Laboratory for Topological Physics, ShanghaiTech University, Shanghai 201210, China}

\author{Yichen Yang}
\affiliation{Center for Excellence in Superconducting Electronics, State Key Laboratory of Functional Materials for Informatics, Shanghai Institute of Microsystem and Information Technology, Chinese Academy of Sciences, Shanghai 200050, China}

\author{Zhengtai Liu}
\affiliation{Center for Excellence in Superconducting Electronics, State Key Laboratory of Functional Materials for Informatics, Shanghai Institute of Microsystem and Information Technology, Chinese Academy of Sciences, Shanghai 200050, China}

\author{Zhe Huang}
\affiliation{Center for Excellence in Superconducting Electronics, State Key Laboratory of Functional Materials for Informatics, Shanghai Institute of Microsystem and Information Technology, Chinese Academy of Sciences, Shanghai 200050, China}

\author{Zhicheng Jiang}
\affiliation{Center for Excellence in Superconducting Electronics, State Key Laboratory of Functional Materials for Informatics, Shanghai Institute of Microsystem and Information Technology, Chinese Academy of Sciences, Shanghai 200050, China}

\author{Xiangle Lu}
\affiliation{Center for Excellence in Superconducting Electronics, State Key Laboratory of Functional Materials for Informatics, Shanghai Institute of Microsystem and Information Technology, Chinese Academy of Sciences, Shanghai 200050, China}
\affiliation{Center of Materials Science and Optoelectronics Engineering, University of Chinese Academy of Sciences, Beijing 100049, China}

\author{Jishan Liu}
\affiliation{Center for Excellence in Superconducting Electronics, State Key Laboratory of Functional Materials for Informatics, Shanghai Institute of Microsystem and Information Technology, Chinese Academy of Sciences, Shanghai 200050, China}
\affiliation{Center of Materials Science and Optoelectronics Engineering, University of Chinese Academy of Sciences, Beijing 100049, China}

\author{Zhonghao Liu}
\affiliation{Center for Excellence in Superconducting Electronics, State Key Laboratory of Functional Materials for Informatics, Shanghai Institute of Microsystem and Information Technology, Chinese Academy of Sciences, Shanghai 200050, China}
\affiliation{Center of Materials Science and Optoelectronics Engineering, University of Chinese Academy of Sciences, Beijing 100049, China}

\author{Jinfeng Jia}
\affiliation{Key Laboratory of Artificial Structures and Quantum Control (Ministry of Education), Shenyang National Laboratory for Materials Science, School of Physics and Astronomy, Shanghai Jiao Tong University, Shanghai 200240, China}

\author{Yanfeng Guo}
\email{guoyf@shanghaitech.edu.cn}
\affiliation{School of Physical Science and Technology, ShanghaiTech University, Shanghai 201210, China}

\author{Jianpeng Liu}
\email{liujp@shanghaitech.edu.cn}
\affiliation{School of Physical Science and Technology, ShanghaiTech University, Shanghai 201210, China}
\affiliation{ShanghaiTech Laboratory for Topological Physics, ShanghaiTech University, Shanghai 201210, China}

\author{Dawei Shen}
\email{dwshen@mail.sim.ac.cn}
\affiliation{Center for Excellence in Superconducting Electronics, State Key Laboratory of Functional Materials for Informatics, Shanghai Institute of Microsystem and Information Technology, Chinese Academy of Sciences, Shanghai 200050, China}
\affiliation{Center of Materials Science and Optoelectronics Engineering, University of Chinese Academy of Sciences, Beijing 100049, China}

\begin{abstract}

Quantum materials with layered kagome structures have drawn considerable attention due to their unique lattice geometry, which gives rise to flat bands co-existing with Dirac-like dispersions. The interplay between strong Coulomb correlations and nontrivial band topology in these systems results in various exotic phenomena. 
Recently, vanadium-based materials with layered kagome structures are discovered to be topological metals, which exhibit charge density wave (CDW) properties, significant anomalous Hall effect, and unusual superconductivity at low temperatures.  
Here we exploit high-resolution angle-resolved photoemission spectroscopy to investigate the electronic structure evolution induced by the CDW transition in a vanadium-based kagome material RbV$_3$Sb$_5$. A remarkable band renormalization in the CDW state is observed, which is consistent with first principles calculations based on an inverse star-of-David superstructure. The CDW phase transition gives rise to a partial energy gap opening at the Fermi level, a shift in the band dispersion, and most importantly, the emergence of new van Hove singularities associated with large density of states, which are absent in the normal phase and may be related to superconductivity observed at lower temperatures. 
Our work would shed light on the microscopic mechanisms for the formation of the CDW and superconducting states in these topological kagome metals.

\end{abstract}

\maketitle


A kagome lattice, which is composed of a two-dimensional network of corner-sharing triangles, provides a unique playground for electrons' correlations and nontrivial band topology. For example, spin systems with antiferromagnetic intersite coupling on a kagome lattice are proposed as promising candidates for realizing quantum spin-liquid states ~\cite{1,2,3,4,5,6}.
Besides, kagome metals usually host electronic structures with  co-existing dispersionless flat bands, Dirac cones, and van Hove singularity (VHS) due to the unique crystal structure, which results in topologically nontrivial states ~\cite{7,N_FeSn,NP_nagtiveMR,S_CSSarpes,SA_AHE,Ar_CDWoptical,Ar_theoVHS,Ar_ARPES_TSS,Ar_MBS}. 

Depending on electron's filling number, a variety of novel quantum states have been theoretically proposed to exist in the kagome lattice systems, including charge fractionalization~\cite{9,10}, density wave orders~\cite{11,12,13,Ar_theoCDW}, topological Chern insulators~\cite{14}, and superconductivity~\cite{13,15} etc.  In particular, when the band filling of a kagome metal is near the VHS, calculations show that several saddle points in the band dispersions emerge near the Fermi level ($E_F$)~\cite{13,32,Ar_CDWoptical,Ar_theoVHS}. 
Coulomb scatterings of electrons between these saddle points would give rise to novel Fermi surface instabilities, which may lead to spin or charge density wave (CDW) states~\cite{28,29,30}.

Recently, a new family of materials AV$_3$Sb$_5$ (A = K, Rb, Cs) with layered kagome lattice structures have been discovered~\cite{PM_ortiz,17,PL_ortiz,SA_AHE,CPS_RB_SCCDW,Ar_STM,23,25,26,27,Ar_CDWoptical,Ar_ARPES_sato,Ar_ARPES_TSS,Ar_ARPES_CDWgap,Ar_MBS}. Previous first principles calculations based on density functional theory (DFT) suggest that the energy bands near the Fermi surfaces in this family of materials are topologically nontrivial characterized by odd $\mathbb{Z}_2$ indices~\cite{17,PL_ortiz}. Moreover, these vanadium-based kagome topological metals undergo CDW phase transitions with the critical temperatures $T_{\rm CDW}$ $\approx$ 78 - 103\,K, below which a new $2\times 2$ inverse star-of-David (ISD) superlattice structure becomes energetically favored ~\cite{Ar_STM,26,Ar_CDWfree,tan2021charge}, leading to a reduction of density of states near $E_F$~\cite{Ar_STM,23,25,tan2021charge}. 
Superconductivity emerges from the CDW states if the temperature is further decreased, and the superconducting transition temperature $T_c\sim 2.5\,$K~\cite{PL_ortiz}. Surprisingly, remarkable anomalous Hall effect (AHE) is observed in the CDW states of these systems~\cite{SA_AHE,22}, although so far no clear evidence of long-range magnetic order has been reported. Despite some early theoretical attempts \cite{Feng_2021,park2021electronic,Ar_theoVHS,lin2021complex,denner2021analysis,Ar_CDWfree}, the microscopic mechanisms behind these intriguing phenomena, such as the driving force for the CDW states, the pairing mechanism and symmetry of the superconducting states, and the origin of AHE, are still open questions.

\begin{figure*}[t]
\includegraphics[width=15cm]{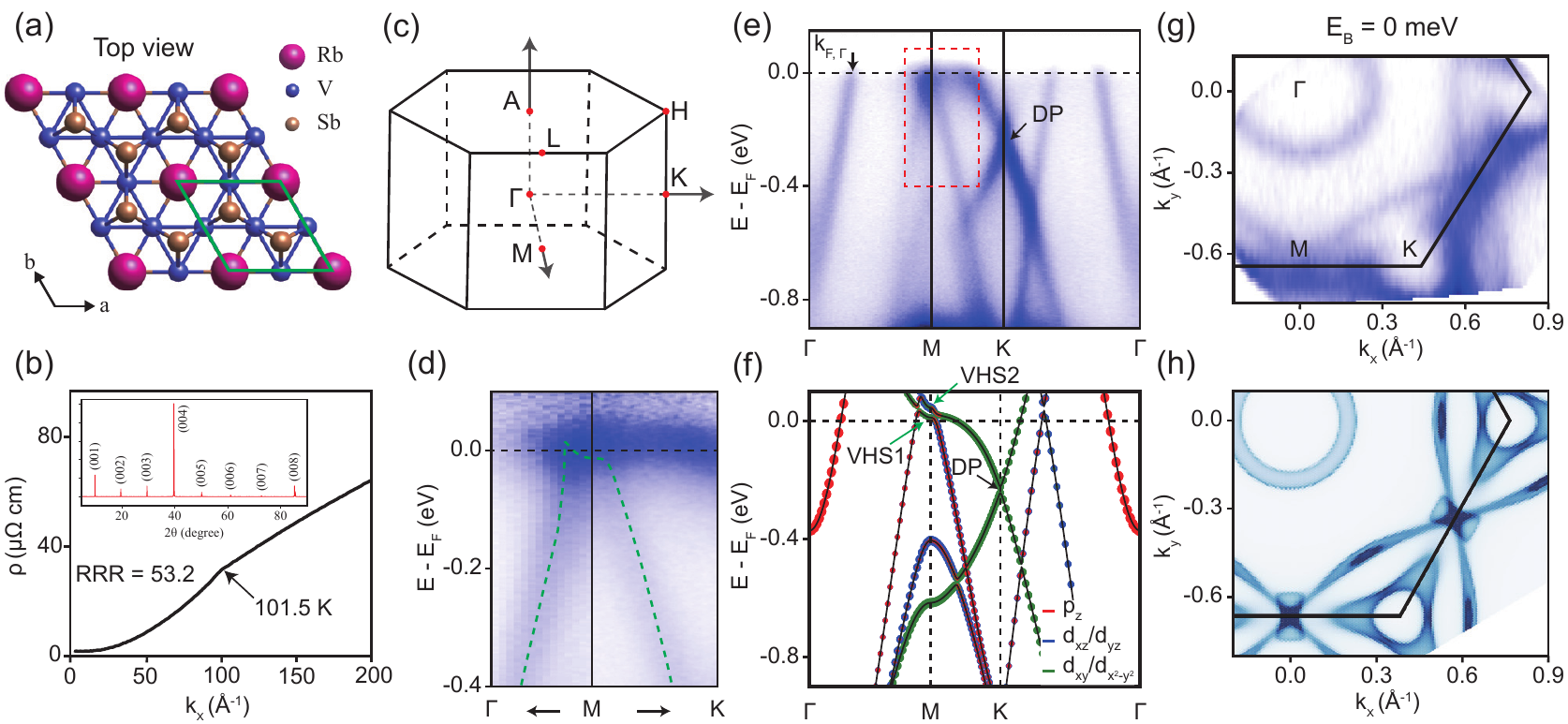}
\caption{(a) The crystal structure of RbV$_3$Sb$_5$ from top view. (b) The resistivity versus  temperature curve with the anomaly near the 101\,K, signifying the charge density wave phase transition. (c) Three-dimensional hexagonal Brillouin zone of the layered kagome lattice. (d) Zoom-in plot of the band dispersion in the red box on (g) near the $M$ point. The ARPES spectra are divided by the Fermi-Dirac distribution function to visualize the band dispersion above $E_F$. The green dotted line is guide for the eye. (e) Experimentally obtained Fermi surface map at 19 eV and above $T_{\rm CDW}$. The back line indicates the hexagonal BZ of the kagome lattice. (f) The energy constant map taken from the band calculations corresponding to the Fermi level of ARPES data.(g) The ARPES spectra above $T_{\rm CDW}$ along the $\Gamma$-$M$-$K$-$\Gamma$ direction and (h) the band calculations with the spin-orbital coupling along the same direction as (g).}
\label{fig1}
\end{figure*}

In this work, combining angle-resolved photoemission spectroscopy (ARPES) and first principles calculations, we study the electronic structure and the microscopic mechanism for the CDW state in RbV$_3$Sb$_5$, which exhibits the highest CDW transition temperature among the AV$_3$Sb$_5$ (A = K, Rb, Cs) materials~\cite{PM_ortiz,PL_ortiz,CPS_RB_SCCDW}. In particular, we have performed systematic measurements for the temperature evolution of the electronic structure in RbV$_3$Sb$_5$. For the first time, we have observed strong band renormalization effects at low temperatures $T<T_{\rm CDW}$. As a result of the CDW transition, we observe a partial energy gap opening at $E_F$ and a shift in the band dispersion to higher binding energy. More interestingly, new VHSs around $M$ points emerge in the CDW phase, which were completely absent in the normal phase. The new VHSs contribute to large density of states near $E_F$, which may be related to superconductivity observed at lower temperatures.  

In addition, by measuring the zero-frequency joint density of states at the Fermi surface, we find significant charge fluctuations at wavevectors connecting two of the $M$ points. We further calculate the phonon dispersions in the normal phase, and find unstable modes at both $M$ and $L$ points in the Brillouin zone. These results imply that both the charge fluctuations at the Fermi surface and the phonon instabilities in the pristine structure are important in driving the system into the CDW state.

RbV$_3$Sb$_5$ crystallizes in a layered hexagonal lattice structure with space group $P6/mmm$ (No. 191). As shown in Fig.~1(a), the system consists of vanadium atoms forming layers of kagome lattices, which are intercalated by Sb atoms forming honeycomb lattice. The vanadium kagome planes are separated by layers of alkali Rb ions forming  triangular networks.
The in-plane hexagonal lattice constant $a$ is 5.472~${\AA}$, and the out-of-plane lattice constant $c$ is 9.073~${\AA}$.
In Fig.~1(b) we show the temperature dependence of the in-plane resistivity for RbV$_3$Sb$_5$, which exhibits an anomaly around 101\,K. This anomaly in resistivity is consistent with previous reports~\cite{PL_ortiz,CPS_RB_SCCDW}, and is believed to be a hallmark of the CDW phase transition.

\begin{figure*}[t]
\includegraphics[width=15cm]{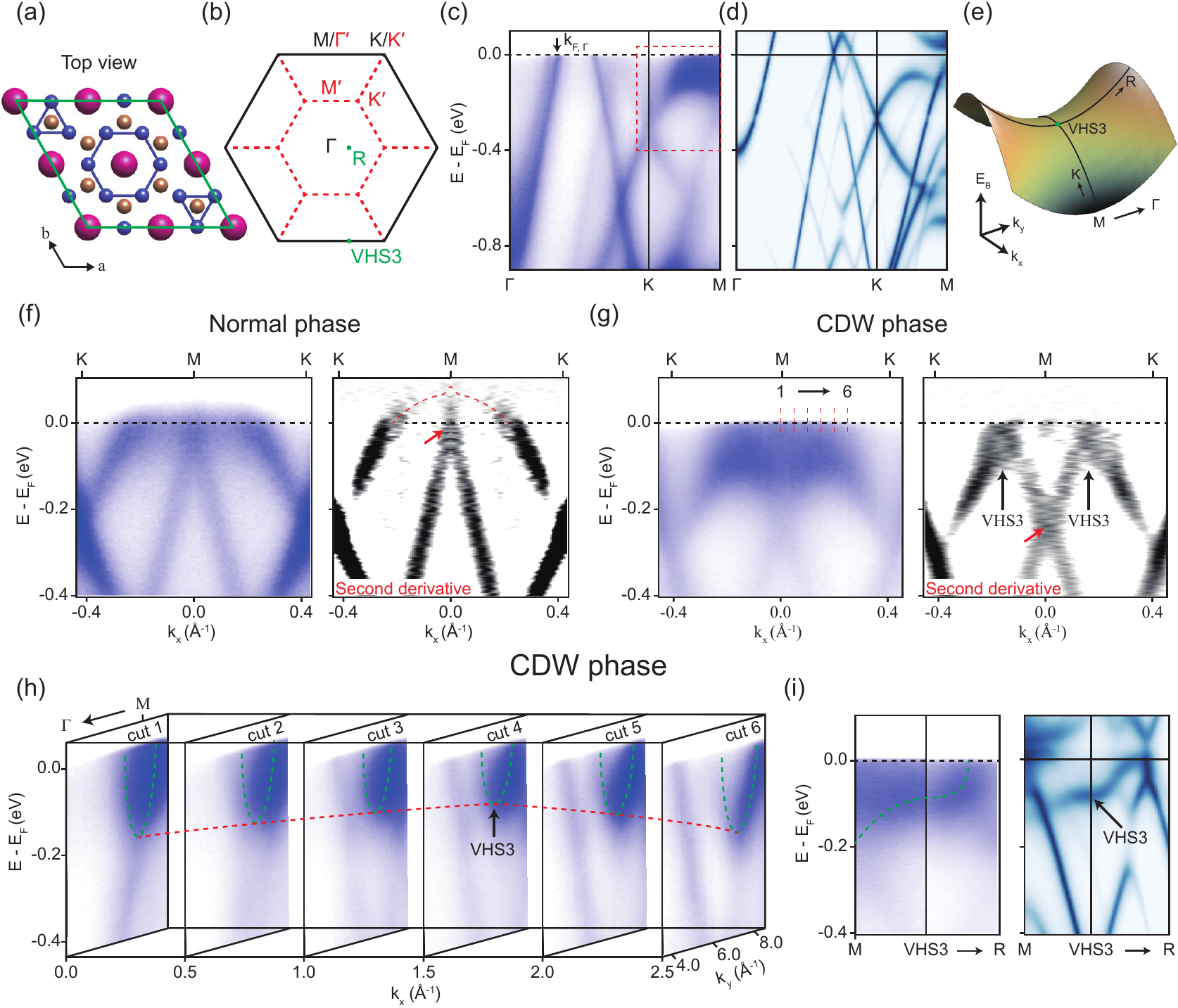}
\caption{(a) The image of vanadium-based kagome lattice under the 2 $\times$ 2 charge density wave, which is consistent with the ISD superlattice. The green line indicates the unit cell of 2 $\times$ 2 superstructure of ISD. (b) The hexagonal Brillouin zone (black line) and the folded Brillouin zone of 2 $\times$ 2 superstructure (red dotted line). (c) ARPES spectra of CDW phase along the $\Gamma$-$K$-$M$ direction taken with 55 eV photon energy at 10 K, below $T_{\rm CDW}$, and (d) the calculated unfolded spectral function for the ISD CDW phase. (e) Schematic picture of the saddle point near the VHS3. The high symmetry cut along the $M$ to $K$ direction obtained from the normal phase ($>$ $T_{\rm CDW}$) (f) and the CDW phase ($<$ $T_{\rm CDW}$) (g). The second derivative spectra of the left panel to enhanced the visibility of the band near the $M$ point. (h) The cut data for the vertical direction of $M$-$K$ direction from cut 1 to 6 as marked in the (g). (i) The ARPES spectra and the calculated unfolded spectra of 2 $\times$ 2 $\times$ 1 ISD along the $M$-$VHS3$-$R$ direction. The green and red dotted line is guide for the eye.}
\label{fig2}
\end{figure*}

We first focus on the normal phase of RbV$_3$Sb$_5$ ($T>T_{\rm CDW}$). According to the detailed photon energy dependent ARPES measurements, we found that the inner potential $V_0\!=\!15.6\,$eV, based on which the value of $k_z$, as well as high-symmetry planes and points in the Brillouin zone (BZ) [Fig.~1(c)] can be precisely determined. Figure~1(e) illustrates the photoemission intensity map in the $k_x$-$k_y$ plane measured at 200\,K with 19\,eV photons, which corresponds to Fermi surface of RbV$_3$Sb$_5$ on the $\Gamma-M-K$ plane. Besides, in Fig.~1(g) we show the photoemission intensity plot along the high-symmetry path connecting $\Gamma$-$K$-$M$-$\Gamma$ points. Combining both Fig.~1(e) and 1(g), it can be clearly seen that the system possesses an electron pocket around $\Gamma$, a Dirac-cone-like dispersion around $K$ point and some relatively flat dispersions with large intensity near $M$ point. 

\begin{figure*}[t]
\includegraphics[width=16cm]{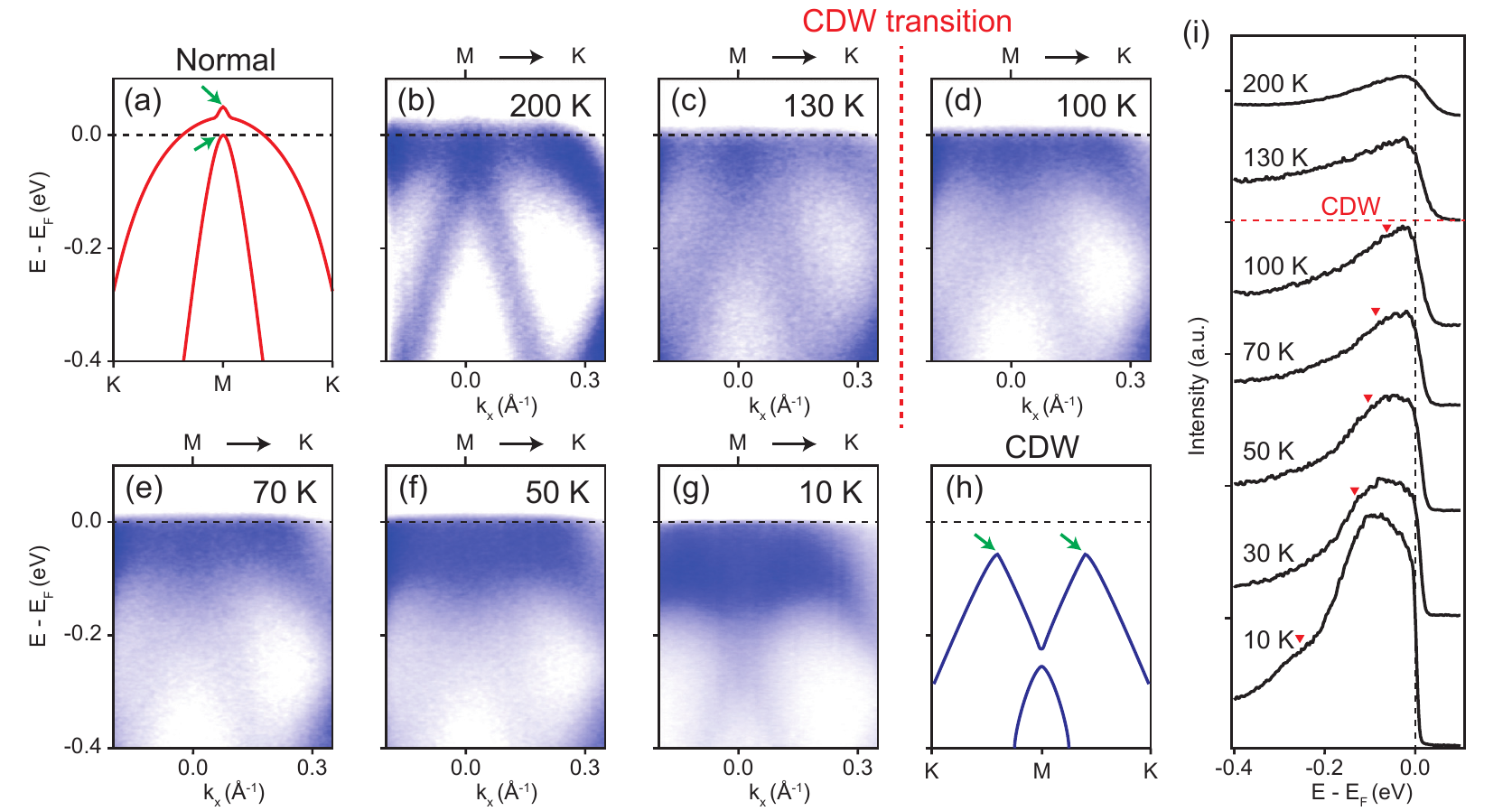}
\caption{(Color online). Temperature-dependent ARPES measurements taken from the 19 eV near the $M$ point. (a) The hole-like band at $M$ point is shifted to higher binding energy as decreasing the temperature. The green arrows indicate the VHS at the schematic figure. (b) Corresponding EDCs at the $M$ point for data in (a) to reveal the evolution of the peak shift. The red mark indicates the emergence of the peak below the CDW transition temperature.}
\label{fig3}
\end{figure*}

In Figs. 1(f) and 1(h) we demonstrate the calculated Fermi surface map on the $\Gamma-M-K$ plane and calculated band structure along the $\Gamma$-$K$-$M$-$\Gamma$ path, respectively, both of which are in good agreement with experiments. Neglecting spin-orbit coupling (SOC), calculations show that the the Dirac point at $K$ point is fourfold degenerate (including spin degeneracy), which is protected by $C_{3z}$, inversion, and time-reversal symmetries of RbV$_3$Sb$_5$ in its normal state. Similar Dirac points in the absence of SOC have been observed in other transition-metal based kagome lattices~\cite{N_FeSn,S_CSSarpes}. Including SOC would open a gap $\sim$20 meV at the Dirac point, which is indeed observed in our ARPES measurement. The calculations further indicate that the Dirac cone near the $K$ point is mostly consisted of vanadium $d_{xy}$ and $d_{x^2-y^2}$ orbitals, while the electron pocket around $\Gamma$ is mainly contributed by the antimony $p_z$ orbitals. Moreover, both ARPES measurements and DFT calculations indicate that the $E_F$ intersects with two VHSs near $M$ point [marked as VHS1 and VHS2 in Fig.~1(h)], which exhibit upward dispersion along $M-\Gamma$ path while downward dispersion along $M-K$ path, as shown in Fig.~1(d). These lead to large density of states around $M$ points near $E_F$, which may introduce Fermi surface instabilities at wavevectors connecting two $M$ points.

We then turn to discuss the electronic structure in the CDW phase of RbV$_3$Sb$_5$. When $T<T_{\rm CDW}$, the system forms a $2\times 2$ ISD type superstructure as shown in Fig.~2(a)~\cite{Ar_STM,tan2021charge,Ar_CDWfree}, and our first principles calculations indicate that such a structure is energetically favored. It is worthy to note  that recent studies suggest there are additional cell doubling or even quadrupling along the $c$ axis~\cite{23,uykur2021optical,park2021electronic}. However, since the states around the VHSs near $M$ points and the Dirac point are of in-plane orbital characters [Fig.~1(f)], 
the cell doubling/quadrupling along the $c$ axis does not significantly change the electronic structure around the $M$ point and the Dirac point, thus in this work we only consider the in-plane 2~$\times$~2 ISD structure in the calculations. The BZ of the ISD supercell is one quarter of the pristine BZ [Fig.~2(b)], therefore the energy bands from the pristine lattice would be folded from the original BZ to the ISD supercell BZ, which may open up gaps at the supercell BZ boundary. However, experimentally we hardly discover any spectral weight of the folded bands as shown in Fig.~2(c), suggesting a mild charge modulation in the CDW phase of RbV$_3$Sb$_5$ with ISD structure.

Although the band-folding effect is weak, we indeed find that the CDW transition would result in an orbital-selective variation of the band structure compared to that of the normal phase. First, as shown in Fig.~2(c), the Fermi wavevector $k_F$ of the electron pocket around $\Gamma$ increases from 0.257\,${\AA}^{-1}$ in the normal state to 0.262\,${\AA}^{-1}$ with a dramatic sinking of this band due to the CDW transition. As the electron pocket around $\Gamma$ is mostly contributed by the out-of-plane $p_z$ orbitals, a strong energy shift at $\Gamma$ implies a possible cell doubling or quadrupling along the $c$ axis. In sharp contrast, the Dirac band around the $K$ point, mainly of in-plane orbital character, remains nearly intact upon the phase transition. Such behavior is reminiscent of that in bulk 1$T$-TiSe$_2$, which is well known as a 2 $\times$ 2 $\times$ 2 CDW material~\cite{PRB_jahn}. This finding implies the possible charge modulation along the \emph{c}-axis in RbV$_3$Sb$_5$, as also suggested by some previous reports~\cite{23,uykur2021optical}.

Besides, the CDW phase transition gives rise to significant change of spectra around $M$ point, as shown by the direct comparison of the photoemission intensity plots measured in the normal and CDW states [Figs. 2(f-g)]. Evidently, nearly the whole VHS1 in the normal state is pushed downwards in binding energy, while only a small region of the higher VHS2 close to the $M$ point is gapped, resulting in the `M'-shaped band dispersion along $K$-$M$-$K$ path in Fig.~2(g). Here second derivative spectra are  provided as well in order to enhance the visibility of the CDW-induced spectral weight shift around the $M$ point. Comparing the energy positions of the VHS1 saddle point in the normal phase and in the CDW phase [indicated by the red arrows in Figs.~2(f-g)], we estimate an energy shift of about 200\,meV induced by the CDW transition. More interestingly, a new VHS (marked as VHS3 thereafter) emerges in the CDW phase, and its saddle point is located somewhere close to $M$, as illustrated in Fig. 2(g). Our experimental finding is fully supported by first principles calculations based on a $2 \times 2\times 1$ ISD structure, as shown by the calculated unfolded spectral function in Fig.~2(d) and in the right panel of Fig.~2(i). Taking into account of the large density of states introduced by this emergent VHS3, it should be intimately relevant to the superconductivity in RbV$_3$Sb$_5$ observed at lower temperatures.

Figs. 3(b-g) show the detailed temperature dependent evolution of VHSs around $M$ across $T_{\rm CDW}$. When the temperature is above $T_{\rm CDW}$, as schematically illustrated in Fig. 3(a), both VHS1 and VHS2 around the $M$ point are in the vicinity of $E_F$, and they almost remain intact with decreasing temperature as shown in Figs.~3(b)-(d). However, once the temperature is below $T_{\rm CDW}$, states around $M$ point are gradually shifted to higher binding energy, eventually leading to a large downshift $\sim\!$ 200 \,meV for states at $M$, and a `M'-shaped band dispersion appears at 10 K [Fig. 3(h)]. Such temperature dependence of the gapped spectral weight around $M$ has unambiguously illustrated its CDW origin. Moreover, we also present the photoemission energy distribution curves (EDCs) taken around $M$ in Fig. 3(i). We found that one more shoulder-like peak (marked by red triangles) in the EDC suddenly emerges below $T_{\rm CDW}$, and is continuously pushed to higher binding energy with the decrease of temperature, which demonstrates a complete picture of the CDW gap development in RbV$_3$Sb$_5$.

In order to better understand the microscopic mechanism for the formation of the CDW state, we calculate the zero-frequency joint density of states (DOS) contributed by the Fermi surface:
\begin{equation}
    C(\mathbf{q})=\frac{\Omega{_0}}{(2\pi)^3}\int d\mathbf{k} \,A(\mathbf{k},E_F) \,A(\mathbf{k}+\mathbf{q},E_F)
\end{equation}, 
where $A(\mathbf{k},E_F)$ is the spectral function at  $E_F$ at $\mathbf{k}$ point in the BZ, and $\Omega_0$ is the volume the primitive cell. 
$C(\mathbf{q})$ describes the phase space  for the scatterings of electrons from the states at $\mathbf{k}$ to those at $\mathbf{k}+\mathbf{q}$ at the Fermi surface, which is a characterization of the Fermi-surface charge fluctuations with wavevector $\mathbf{q}$.
Therefore, it is  expected that $C(\mathbf{q})$ would show a peak at the corresponding ordering wavevector if there exist any charge instability induced by charge fluctuations at the Fermi surface. 
This autocorrelation of ARPES spectra in the normal state has been demonstrated to give a reasonable count for charge-ordering instabilities of various compounds \cite{shen2007novel,chatterjee2006nondispersive,mcelroy2006elastic,shen2008primary}. The $C(\mathbf{q})$ map extracted from APRES spectra is shown in Fig.~4(a), which exhibit several peaks at wavevectors $\mathbf{q}_{i}$ (i=1, 2 and 3). This is consistent with the calculated zero-frequency joint DOS as shown in Fig.~4(b). We find that $\{\mathbf{q}_{i}\}$  are exactly the ``nesting" wavevectors connecting the different saddle points (around $M$) at the Fermi surface in the BZ, as schematically shown in Fig. 4(c).  This indicates that the strong charge fluctuations due to the Fermi-surface nesting is an important driving force for the formation of CDW state in RbV$_3$Sb$_5$.

Moreover, we have also calculated the phonon band structure along the high-symmetry path as shown in Fig.~4(d), which clearly shows two unstable modes around $M$ and $L$ points respectively. The soft phonon mode at $M$ point and the strong charge fluctuations at $\mathbf{q}_1$ would cooperate with each other, both of which tend to drive the system into a CDW state that doubles the primitive cell along the $\mathbf{q}_1$ direction in the kagome plane. However, since there are three equivalent $\mathbf{q}_1$ wavevectors (three equivalent $M$ points) in the BZ, these three $\mathbf{q}_1$ vectors may be linked with the three sublattices in the kagome lattice, forming a "triple-Q" CDW state preserving $C_{3z}$ symmetry~\cite{Feng_2021}, which is exactly the $2 \times 2$ ISD structure. The additional soft phonon mode at $L$ point may be the origin of the cell doubling along the $c$ axis in the CDW phase.

\begin{figure}[t]
\includegraphics[width=8.3cm]{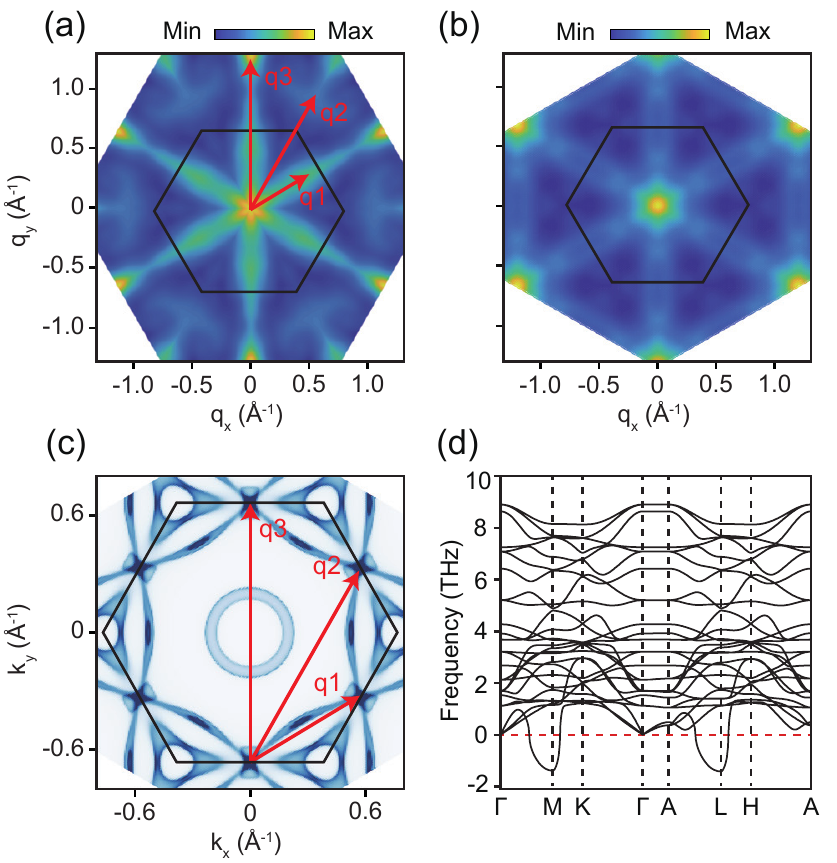}
\caption{(Color online). (a) Two dimensional joint DOS results on the saddle band point regions. (b) Calculated joint DOS at the Fermi energy. (c) Calculated Fermi surface projected to the $k_x$-$k_y$ plane. (d) Calculated phonon spectrum along a high-symmetry path. The high symmetry points are marked in Fig.~1(c).}
\label{fig4}
\end{figure}

To summarize, we have studied the electronic structures of a kagome topological metal RbV$_3$Sb$_5$ both in the normal phase and in the CDW phase.  Our APRES measurements indicate strong band renormalization effects due to the CDW phase transition. These include partial gap opening at $E_F$, orbital-selective energy shift of the dispersion, and most importantly, the emergence of new VHSs near $E_F$ induced by the CDW transition. The band renormalizations around the $K$ and $M$ points, as well as the emergence of new VHSs, can be fully captured by first principles calculations based on a $2 \times 2$ ISD supercell. The new VHSs in the CDW phase contribute to large DOS near $E_F$, which may  be relevant with the emergence of superconductivity at lower temperatures.  We further investigate the microscopic mechanisms for the formation of the CDW state, and find that the system exhibits strong charge fluctuations at wavevectors connecting two adjacent saddle points at the Fermi surface. The Fermi-surface charge fluctuations would cooperate with a soft mode at the same wavevector, eventually driving the system into the  CDW state observed in experiments. Our work is a significant step forward in understanding the mysterious CDW phase in the V-based topological kagome metal systems, and will provide useful guidelines for future experimental and theoretical studies.

\section*{Methods}

\subsection{Single crystal synthesis and characterization.}
The RbV$_3$Sb$_5$ crystals were grown by using a mixture of RbSb$_2$ and RbSb precusors as the flux. The pre-reacted RbSb$_2$, RbSb and VSb$_2$ were mixed in a molar ratio of 2 : 2 : 1, placed into an alumina crucible, and then sealed into a quartz tube. The assembly was heated up to 1000$^{\circ}$C in 10 h in a weel-type furnace, maintained at the temperature for more than 30 h, and then slowly cooled down to 450$^{\circ}$C at a temperature decreasing rate of 1.5$^{\circ}$C/h. Then the furnace was shut down and naturally cooled to room temperature. Crystals with black shiny metallic luster were obtained in the aluminium crucible. The crystallographic phase and quality of the crystals were examined on a Bruker D8 VENTURE powder diffractometer using Cu K$\alpha$ radioactive source ($\lambda$ = 1.5418~$\AA$) at room temperature. During measurement, the crystals were aligned along the (001) plane. The resistivity of the crystals was measured by using a standard four-wire method in a commercial DynaCool physical properties measurement system from Quantum Design.

\subsection{Angle-resolved photoemission spectroscopy.}
ARPES measurements were performed at both 03U beamline of Shanghai Synchrotron Radiation Facility (SSRF)~\cite{yang2021high} and 13U beamline of National Synchrotron Radiation Laboratory (NSRL). These two endstations are both equipped with Scienta-Omicron DA30 electron analyzers. The angular and the energy resolutions were set to 0.2$^\circ$ and 8 $\sim$ 20 meV (dependent on the selected probing photon energy). All samples were cleaved in an ultrahigh vacuum better than 8.0 $\times$ 10$^{-11}$ Torr.

\subsection{First principles calculations of the bulk electronic structure.}
Density functional theory (DFT) calculations are performed with the Vienna ab initio simulation package (VASP) which adopts the projector-augmented wave method \cite{KRESSE199615}. The plane-wave energy cutoff is set at 400 eV. The exchange-correlation functional of the Perdew-Burke-Ernzerhof (PBE) type\cite{PhysRevLett.78.1396,PhysRevB.50.17953,PhysRevB.59.1758} is used for both structural relaxations and self-consistent electronic calculations. The convergence criteria for the electronic iteration and geometry optimization are set to 10$^{-6}$ eV and 0.001 eV/\AA, respectively. The BZ is sampled by a 9$\times$9$\times$7 k-mesh for the normal phase and a 6$\times$6$\times$4 k-mesh for the CDW phase within the Gamma-centered scheme. Phonon spectrum is calculated with PHONOPY code \cite{phonopy}, using density-functional perturbation theory (DFPT) method. Joint DOS is calculated based on Wannier orbitals which are obtained through VASP2WANNIER90 interface \cite{Arash2008}, 60 orbitals are included with SOC being considered.

\section*{Acknowledgements} 

This work was supported by the National Key R$\&$D Program of the MOST of China (Grant No. 2016YFA0300204), the National Science Foundation of China (Grant Nos. U2032208, 92065201, 11874264), and the Natural Science Foundation of Shanghai (Grant No. 14ZR1447600). Y. F. Guo acknowledges the starting grant of ShanghaiTech University and the Program for Professor of Special Appointment (Shanghai Eastern Scholar). Part of this research used Beamline 03U of the Shanghai Synchrotron Radiation Facility, which is supported by ME$^2$ project under contract No. 11227902 from National Natural Science Foundation of China. J. P. Liu acknowledges the start-up grant of ShanghaiTech University. The authors also thank the support from Analytical Instrumentation Center (\#SPST-AIC10112914), SPST, ShanghaiTech University. 



\bibliography{RbVSb}

\end{document}